\documentclass[aps,pra,twocolumn,groupedaddress]{revtex4-1}
\usepackage{amsmath,amssymb,graphicx}

\usepackage{hyperref}
\hypersetup{
    pdftitle={Nonlinear optics at low powers: New mechanism of on-chip optical frequency comb generation},
    pdfauthor={Andrei Rogov},
    pdfsubject={Nonlinear Optics},
    pdfkeywords={Nonlinear optics, Integrated optics, Resonators, Silicon photonics, Frequency combs, Instabilities and chaos},
    bookmarksnumbered=true,
    bookmarksopen=true,
    bookmarksopenlevel=1,
    colorlinks=true,
    linkcolor=blue,
    citecolor=blue,
    urlcolor=black,
    allcolors=black,
    pdfstartview=Fit,
    pdfpagemode=UseOutlines,    % this is the option you were lookin for
    pdfpagelayout=TwoPageRight
}

\begin{document}

%\preprint{}

\title{Nonlinear optics at low powers: \\
New mechanism of on-chip optical frequency comb generation}


\author{Andrei S. Rogov}
\email[]{arogov@purdue.edu}
%\thanks{}
\author{Evgenii E. Narimanov}
\affiliation{Birck Nanotechnology Center, School of Electrical and Computer Engineering, Purdue University, West Lafayette, Indiana 47907, USA}


\date{\today}

\begin{abstract}
Nonlinear optical effects provide a natural way of light manipulation and 
interaction, and form the foundation of applied photonics -- from high-speed 
signal processing and telecommunication,
% \cite{willner2014, salem2008, taeed2005, pelusi2007, radic2008}, 
to ultra-high bandwidth interconnects 
%\cite{alduino2007, miller2000, lee2008, levy2010} 
and information processing.
%\cite{woods2012, caulfield2010, brunner2013}.
However, relatively weak nonlinear response at optical frequencies 
%\cite{boyd2008}
calls for operation at high optical powers, or boosting efficiency of nonlinear 
parametric processes by enhancing local field intensity with high quality-factor 
resonators near cavity resonance, resulting in reduced operational bandwidth and increased 
loss due to multi-photon absorption. We present an alternative to this 
conventional approach, with strong nonlinear optical effects at low 
local intensities, based on period-doubling bifurcations near nonlinear cavity 
anti-resonance, and apply it to low-power optical frequency comb generation 
%\cite{kippenberg2011,torres-company2014} 
in a silicon chip.
\end{abstract}

%\keywords{}

\maketitle

\section{Introduction}

In the never-ending quest for higher signal processing speeds, optics-based systems
%With electronics-based computing reaching its theoretical speed limits, replacing electronic components with optical counterparts 
represent an attractive research direction, since, in contrast to the well-studied area of electronics, optical domain is perfectly suitable for operation at high frequencies.
The basis for any signal processing is manipulation of signals, or light waves, as in the case of optics. Medium nonlinearity provides a means of interaction of different propagating waves with each other and the medium itself.

Two strategies have been traditionally used to enhance nonlinear optical effects: (1) targeting materials with high optical nonlinearities, such as chalcogenide glasses \cite{taeed2007, yeom2008}, silicon \cite{salem2008, leuthold2010}, $AlGaAs$ \cite{aitchison1997, van2002-1},
and (2) employing resonant structures to increase local field intensity. % \cite{•}. 
The choice of material is oftentimes dictated by fabrication limitations and the overall design compatibility requirements. For example, due to its lower cost and direct compatibility with the well developed CMOS industry, silicon-based nonlinear photonics has gained a lot of interest in the last decade \cite{leuthold2010}.
On the other hand, resonators offer significant enhancement of local field intensity at the resonance regime,
which allowed for experimental observation of nonlinear optical effects in mode-locked lasers 
\cite{spence1991, brabec1992}
and fiber-ring resonators at first
%Nonlinear optical effects were observed in fiber-ring cavities at first 
\cite{shelby1988, nakazawa1988, coen2001},
and then, with the development of microfabrication, in high quality-factor ($Q$) microresonators 
\cite{kippenberg2004, delhaye2007, 
agha2009, 
liang2011, herr2012, 
papp2011, 
li2012, 
razzari2010, 
hausmann2014, 
jung2013, 
levy2010, miller2015, 
ferdous2011, liu2014, xue2015, xue2015-1, 
griffith2015}.
However, boosting local field intensities with high-$Q$ resonators for enhancing nonlinearity has its disadvantages.
First, high intracavity intensities lead to significant multi-photon absorption losses, which makes this resonant approach inapplicable to the materials with substantial nonlinear losses, such as silicon at the telecom wavelength of 1550 nm \cite{lau2015}.
%-- the most widely used wavelength in the modern optical communication systems.
Second, ultra-high-$Q$ microresonators, which allow for observation of nonlinear effects at the lowest power, are extremely sensitive to fabrication non-idealities and suffer from poor scalability and on-chip integrability.
Finally, high-$Q$ microresonators at the resonance are highly susceptible to external perturbations -- up to individual molecules -- which is advantageous for nanoparticle detection systems \cite{armani2007, vollmer2008, zhu2010}, but detrimental to many other applications. As we show in this work, nonlinear effects in microresonators can also be observed in the near-anti-resonance regime, which in contrast to the standard resonant approach, naturally implies low intracavity intensity operation, 
%and, as a consequence, is much less prone to the aforementioned disadvantages.
and as such could be less prone to the aforementioned disadvantages.
Here, the anti-resonant operation of a nonlinear cavity excited by continuous wave (CW) pumping
%, we imply that 
is defined as the case when 
the total phase detuning of the pump frequency from the closest resonant mode of the cavity is equal to
$\pi$: $|\phi + \phi_{NL}| = \pi$, where $\phi$ is the linear contribution 
and $\phi_{NL}$ -- nonlinear (intensity-dependent).
The linear detuning $\phi$ is calculated in the conditions of low intensity, 
such that nonlinear effects are negligible and the cavity is considered ``cold''.
The nonlinear contribution $\phi_{NL}$ is the extra phase accumulated by the intracavity field 
over one roundtrip in the cavity due to the optical Kerr effect.

Effects of nonlinearity on a system dynamics can often be understood by introducing the concept of modulational instability (MI).
%If a stable state of the system corresponds to a uniform carrier wave propagating in a nonlinear dispersive medium, an unstable state can be viewed as a modulated carrier wave. 
MI caused by the interplay between nonlinear and dispersive effects has been observed in many areas of physics \cite{zakharov2009}, including nonlinear optics, where it manifests itself as the breakup of
% continuous wave (CW)
CW radiation into a train of ultra-short pulses \cite{agrawal2001}. 
When a CW beam propagates through a homogeneous nonlinear dispersive optical medium, anomalous dispersion is required for the modulational instability to occur. 
However, in the presence of a feedback, as in the case of a resonator system, modulational instability can arise even at normal dispersion \cite{haelterman1992}, 
and occurs either close to the cavity resonance, or close to the cavity anti-resonance \cite{coen1997}.
Thus, nonlinear effects in resonators should be expected in both the resonance and anti-resonance regime. 
Motivated by the advantages of low intracavity intensity operation,
we are studying the effects of nonlinearity in a resonator system in the vicinity of the cavity anti-resonance, and applying this approach to low-power optical frequency comb generation \cite{kippenberg2011,torres-company2014} in a silicon chip at the telecom wavelength.

% Previous work
Previous work on frequency comb generation has been primarily focused on the resonant regime of a high-$Q$ nonlinear cavity \cite{kippenberg2011,torres-company2014}. In this case the mean-field approximation is valid, and the dynamics can be modeled accurately using either the modal expansion theory \cite{chembo2010-2} or the Lugiato-Lefever equation (LLE) \cite{lugiato1987,haelterman1992,chembo2013,coen2013}.
%These approaches are applicable only if the overall cavity detuning from the resonance, including both the linear and nonlinear contributions, is small.
In the present work we are interested in the anti-resonant regime of a cavity, 
%so that the overall cavity detuning from the resonance is of the order of $\pi$, and 
in which the system exhibits period-doubling behaviour, that has been observed to occur in fiber-ring resonators \cite{nakatsuka1983, vallee1991, steinmeyer1995, coen1998}. 
The mean-field models can not be applied to the kind of period-doubling instabilities, which origin lies in the inhomogeneity of the pump field and which cause significant changes in the intracavity field between consecutive roundtrips.
%since the latter occur on the time-scale of multiple roundtrips, 
For that reason, we resort to using the Ikeda map \cite{ikeda1979,ikeda1980} and solving numerically the nonlinear Schr\"odinger's (NLS) equation \cite{agrawal2001,yin2007,lin2007}.

\section{Results}

\subsection{New mechanism (single resonator system)}
In the anti-resonant regime, the nonlinear dynamics of the single resonator system is affected by period-doubling instabilities \cite{coen1997}: 
an increased effective nonlinearity in the resonator brakes the system integrability, which results in period-doubling transition to chaos \cite{ikeda1979,ikeda1980}.
If CW-states are considered stable period-1 states, period-doubling bifurcations lead to the formation of regions in the system parameter space with stable states of higher periods -- 
states with multiple wave amplitude values.
Existence of higher-period stable states leads to switching in time-domain: the system traverses the set of allowed stable wave amplitude values in sequence switching between them within the system characteristic time period (roundtrip time in the resonator). Periodic switching in time-domain translates to the comb spectrum in frequency domain.
Thereby, frequency comb generation can be achieved in the anti-resonant regime of the cavity, owing to the existence of period-doubling instabilities.

As we pointed out earlier, a cavity is in the anti-resonant regime if the overall phase detuning $|\phi + \phi_{NL}|$ of the pump frequency from the cavity nearest resonance mode is equal to $\pi$. Therefore, there are two ways to reach the anti-resonant regime.
In the first approach, the pump is tuned into one of the ``cold cavity'' resonances (linear phase detuning $\phi \approx 0$). Then, the nonlinear phase shift $\phi_{NL}$ caused by the increased intracavity intensity pushes the cavity into anti-resonance. This approach requires significant nonlinear phase shift to be accumulated per every roundtrip in the cavity, thus leading to high intracavity intensities ($\phi_{NL} \approx \gamma{L}{P} \approx \pi$) necessary to reach anti-resonance and period-doubling bifurcations \cite{rogov2014}.

On the other hand, if the pump was initially tuned into one of the ``cold cavity'' anti-resonances (linear phase detuning $\phi \approx \pi$), little nonlinear phase shift and low intracavity intensity ($\phi_{NL} \approx \gamma{L}{P} \ll \pi$) is required for the system to reach the anti-resonant regime.
As a result, period-doubling bifurcations, and frequency comb generation in particular, in principle, could be achieved at close to zero intracavity intensities. For real applications low intracavity power means low multi-photon absorption losses. This is the key idea behind the new mechanism we propose.

\begin{figure}[htbp]
\centering
\includegraphics[width=1.0 \columnwidth]{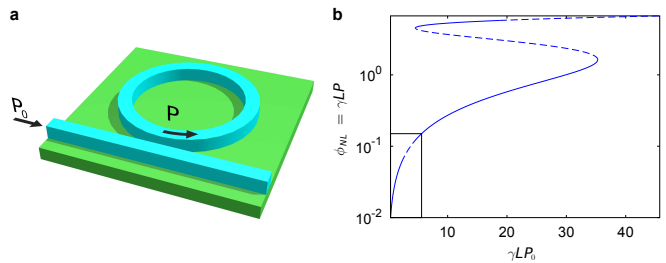}%\columnwidth
\caption{\label{fig:fig1} (a) Schematics of the microring resonator system. (b) Period-1 stable (solid line) and unstable (dashed) states of the intracavity power $P$ at different values of the input power $P_0$ with the highlighted first instability region (black rectangle contour).
The simulation data were obtained for a resonator with ring radius $R=10\ \mathrm{{\mu}{m}}$, $\kappa^2 = 0.1$, $\phi = \pi-0.14$, $\alpha=0.7\ \mathrm{dB/cm}$, and TPA as in silicon near $\lambda=1550\ \mathrm{nm}$.}
\end{figure}

%% Model
In order to illustrate the main features of the suggested mechanism, we first consider a single resonator system without group velocity dispersion (GVD), with a microring resonator coupled to the waveguide (Fig.\ref{fig:fig1}a) used both to pump the resonator and to direct the output signal.
%Zero GVD allows us to apply a simpler, map-based approach for the analysis of the system.
Dispersion-free system allows us to apply a simpler, map-based approach for its analysis.
Then we perform a full spatiotemporal numerical modeling of the nonlinear dispersive system to study the effects of finite GVD on the proposed mechanism.

Throughout the analysis, we take two-photon absorption (TPA) into consideration. However, it is neither essential nor beneficial for our approach to apply. The motivation behind including TPA into the model is to provide a quantitative connection to the experiments with materials where TPA is present or cannot be neglected, such as silicon at 1550 nm.

The wave propagation over one roundtrip of the resonator, assuming a single spatial mode, is governed by the well-known NLS equation \cite{agrawal2001,yin2007,lin2007}
\begin{equation}
\frac{\partial{A}}{\partial{z}} =
i \gamma(1+ir){|A|}^2 A - \frac{i \beta_2}{2}\frac{\partial^2{A}}{\partial{T^2}} - \frac{\alpha}{2} A,
\label{eq:nls}
\end{equation}
where $A = A(z,T)$ is the normalized wave packet amplitude ($|A|^2$ has the units of power), $\gamma$ is the nonlinearity coefficient, $r$ is the TPA coefficient ($r \approx 0.1$ for silicon at 1550 nm), $\beta_2$ is the group velocity dispersion coefficient, $\alpha$ is the power attenuation constant, and $T \equiv  t - z/v_g$ is time in the frame of reference moving with the wave packet along the circumference of the ring at the group velocity $v_g$.
Without GVD, Eq.~(\ref{eq:nls}) is reduced to
\begin{equation}
\frac{\partial{A}}{\partial{z}} =
i \gamma(1+ir){|A|}^2 A - \frac{\alpha}{2} A,
\label{eq:nls-nogvd}
\end{equation}
and has an analytic solution \cite{yin2007}
\begin{eqnarray}
A(L,T) &=& A(0,T)\frac{\exp\left(-\frac{\alpha}{2}L\right)}{\sqrt{1+2r\gamma\tilde{L}{|A(0,T)|}^2}}\nonumber\\
&&\times
\exp\left[{i\ln\left(1+2r\gamma\tilde{L}{|A(0,T)|}^2\right)}/(2r)\right],
\label{eq:nls-nogvd-sol}
\end{eqnarray}
where
\begin{equation}
\tilde{L} = \frac{1-\exp\left(-{\alpha}L\right)}{\alpha}.
\label{eq:nls-nogvd-Leff}
\end{equation}

The coupling of the resonator to the waveguide can be modeled by the matrix equation \cite{yariv2000}:
\begin{equation}
\left[ \begin{array}{c}
b(t) \\
d(t)
\end{array} \right] = 
\left[ \begin{array}{cc}
\tau & \kappa \\
-\kappa^* & \tau^*
\end{array} \right] 
\left[ \begin{array}{c}
a(t) \\
c(t)
\end{array} \right],
\label{eq:coupler}
\end{equation}
where $\tau$ and $\kappa$ are respectively the amplitude transmission and coupling coefficients for the coupler between the microring and the waveguide, $a \equiv A(z = L)$ ($L$ is the circumference of the resonator) and $b\equiv A(z = 0)$, while the amplitudes $c$ and $d$ correspond to the field amplitudes at the ``input'' and ``through'' ports of the waveguide. 
The coupling matrix is unitary so that ${|\tau|}^2 + {|\kappa|}^2 = 1$.
For the single resonator system (Fig.\ref{fig:fig1}a), the CW driving field $c(t) \equiv c = {\rm const}$ from the pump is coherently added through the coupler every roundtrip to the wave circulating in the ring.
From Eq. (\ref{eq:coupler}) we obtain that the intracavity field $b^{(n+1)}(t)$ at the beginning of $(n+1)$-th roundtrip can be related to the field $a^{(n)}(t)$ at the end of $n$-th roundtrip as
\begin{equation}
b^{(n+1)} = \tau a^{(n)} \exp\left(i\phi\right) + \kappa c,
\label{eq:NLRMapEq1}
\end{equation}
where $\phi$ is 
the linear phase detuning of the pump frequency from the cavity nearest resonant mode.

The evolution of the intracavity field through one roundtrip in the resonator is described by Eq. (\ref{eq:nls-nogvd-sol}) and in terms of roundtrip variables takes the form
\begin{eqnarray}
a^{(n)} &=& b^{(n)}\frac{\exp\left(-\frac{\alpha}{2}L\right)}{\sqrt{1+2r\gamma\tilde{L}{|b^{(n)}|}^2}}\nonumber\\
&&\times
\exp\left[{i\log\left(1+2r\gamma\tilde{L}{|b^{(n)}|}^2\right)}/(2r)\right].
\label{eq:NLRMapEq2}
\end{eqnarray}

%% Map. Stability Analysis
The initial condition (\ref{eq:NLRMapEq1}) together with the intracavity evolution equation  (\ref{eq:NLRMapEq2}) constitute a finite-dimensional Ikeda map \cite{ikeda1979,ikeda1980}, which describes the dynamics of a ring resonator at zero GVD.
Stability analysis of the map reveals that the system has multiple regions of period-1 stable and unstable states (Fig.\ref{fig:fig1}b). 
However, it is the very first instability region that allows for period-1 unstable states to exist at the lowest both intracavity $P$ and input $P_0$ power, which is achieved when the resonator is tuned into the vicinity of anti-resonance. At the point where period-1 state looses stability, the system undergoes a period-doubling bifurcation which leads to the formation of period-2 stable state.
This dynamics is illustrated in Fig.\ref{fig:fig2}a that depicts the lowest power period-2 bubble. The originally stable period-1 mode corresponding to the time-independent power inside the ring (solid blue line in Fig.\ref{fig:fig2}a, eventually loses stability with increasing power and the system switches to a new period-2 stable mode -- a state with the period $2T_c$ ($T_c =  L/v_g$ -- roundtrip time) corresponding to two ring roundtrips ({closed red loop with circle markers in Fig.\ref{fig:fig2}a). 
At this point, the steady-state power in the microring is no longer time-independent and the switching between the two power levels occurs (Fig.\ref{fig:fig2}b), leading to multiple subbands in the power spectrum and frequency comb generation (Fig.\ref{fig:fig2}c).

\begin{figure}[htbp]
\centering
\includegraphics[width=1.0 \columnwidth]{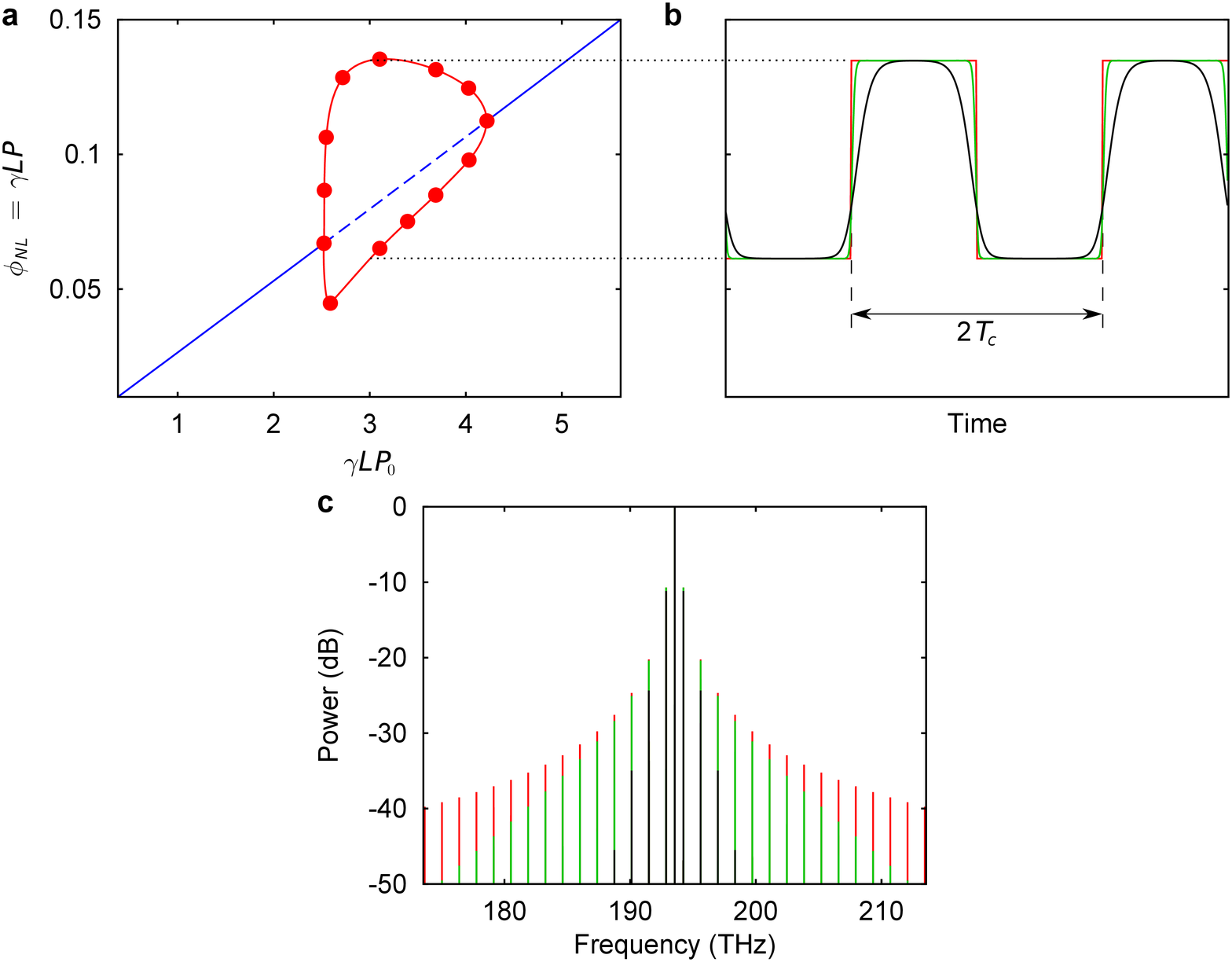}
\caption{\label{fig:fig2} (a) Bifurcation diagram of the first period-1 instability region (black rectangle contour in Fig.\ref{fig:fig1}b) for the states of the intracavity power $P$ at different values of the input power $P_0$.
Fixed points of period-1 and period-2 form the blue line and the red closed loop with circle markers respectively.
Solid lines represent stable fixed points, dashed -- unstable.
(b) Corresponding waveforms and (c) power spectra inside the microring resonator. 
The red (dark gray) curve in (b) and (c) represents the case of zero GVD;
green (light gray) and black curves correspond to the weak $({\beta_2}{L}/{T_c^2} = 1\times10^{-5})$ and 
strong $({\beta_2}{L}/{T_c^2} = 3\times10^{-4})$ normal GVD respectively.
The simulation data were obtained for a resonator excited at $\lambda=1550\ \mathrm{nm}$ ($f \approx 194\ \mathrm{THz}$) with the scaled input power $\gamma{L}{P_0}=3$, $\kappa^2 = 0.1$, $\phi = \pi-0.14$, linear losses $\alpha=0.7\ \mathrm{dB/cm}$, TPA as in silicon near $\lambda=1550\ \mathrm{nm}$, and ring radius $R=10\ \mathrm{{\mu}{m}}$.}
\end{figure}

%% Numerical solution
However, with a finite GVD present in the system, the dynamics of the ring resonator cannot be described by a finite-dimensional map, and the evolution of the intracavity field must be found by integrating the NLS equation (\ref{eq:nls}).
We solve numerically the NLS equation with the initial condition (\ref{eq:NLRMapEq1}) with the FDTD Hopscotch method \cite{taha1984}. 
We seek a numerical solution for $A(z,t)$ at a set of points $z_m, t_k$ on a rectangular grid in the $z, t$ plane, where $z_m = \Delta{z} \cdot m, t_k = \Delta{t} \cdot k$, $\Delta{z}$ is the increment in $z$ and $\Delta{t}$ is the increment in $t$.
The time-step $\Delta{z}$ is chosen large enough to provide a sufficient frequency window for the generated comb spectrum. Step $\Delta{z}$ is adjusted to provide the required accuracy.
The initial condition has a feedback with the time delay equal to the roundtrip time in the resonator. For that reason the numerical integration is performed iteratively, covering $(K-M)$ points in time per pass, where $K$ and $M$ are the number of steps per one roundtrip in t-space and z-space respectively.

Note that the numerical approach of solving the NLS equation we apply in this work is substantially different from the common one (see, e.g., \cite{agha2009}), which is based on iterations with the split-step Fourier method. 
The reason for this is that the discrete Fourier transform with the temporal window span of the single roundtrip time $T_c$ cannot be applied here, since the minimum period of observed states is $2T_c$. Instead, we use the FDTD method \cite{taha1984} to solve numerically the NLS equation with the initial condition defined by the coupling between the waveguide and the resonator.
%Even though FDTD methods are generally slower than split-step methods making use of the fast Fourier transform algorithm \cite{agrawal2001}, and the required computation would take hundreds of millions of roundtrips to reach a steady state in the case of a high-$Q$ resonator \cite{agha2009}, it takes only a few thousand roundtrips for a strongly coupled resonator ($\kappa^2 = 0.1$ in Fig.\ref{fig:fig1}), owing to the much shorter build-up time ($T_b \sim \kappa^{-2}$).

As expected, the numerical solution shows that the instantaneous switching demonstrated earlier without GVD is replaced with smooth transitions (Fig.\ref{fig:fig2}b) in the case of normal GVD.
Dispersion acting together with Kerr nonlinearity effectively limits the frequency comb spectrum: 
the stronger the GVD is, the less frequency components are visible (Fig.\ref{fig:fig2}c).

It should be emphasized that the proposed mechanism of frequency comb formation is qualitatively different from the well known approach 
\cite{kippenberg2004,ferdous2011,herr2012}.
According to the standard picture, all comb sidebands are formed at multiple or single free spectral ranges (FSR) away from the pump 
\cite{herr2012}.
In contrast, in the suggested mechanism, the first harmonic is generated at $f_1 = \frac{FSR}{2} = \frac{1}{2T_c}$ away from the pump frequency, although the higher harmonics follow single FSR spacing (Fig.\ref{fig:fig2}c).

Note that the bifurcation from period-1 to period-2 stable states (threshold point for frequency comb generation) takes place at the nonlinear phase shift per round trip $\phi_{NL}^{Th} \equiv \gamma{L}{P^{Th}} \approx 0.06 \ll 1$ (Fig.\ref{fig:fig2}a).
Moreover, our analysis shows that the threshold intracavity power can be further decreased: 
$\phi_{NL}^{Th} \equiv \gamma{L}{P^{Th}} \sim \kappa^2$, with the optimally tuned linear phase detuning $\phi \sim \pi-\kappa^2$, and the intracavity linear losses negligibly small compared to the nonlinear ones (valid approximation for the system with the parameters given in Fig.\ref{fig:fig2}).
Since the TPA-related loss is proportional to the intracavity power, and other multi-photon absorption processes exhibit even stronger power dependence, nonlinear losses are negligible, and linear losses become the limiting factor for the threshold intracavity power.

However, even though the intracavity power threshold can be low, the input power threshold $P_0^{Th}$ is quite high $\left(\gamma{L}{P_0^{Th}} \approx 2.5\right)$, since the resonator is in anti-resonance regime. Low intracavity power leads to a possibility of using materials with high nonlinear losses, such as silicon at the telecom wavelength, while high input power prohibits the single resonator design discussed above from the practical implementation on a chip. In the next section we resolve this issue.

\subsection{Double resonator system: a solution to the high input power problem}
%% Model & Map
As pointed out earlier, high threshold input power $\left(\gamma{L}{P_0^{Th}} \approx 2.5\right)$ 
of the single resonator system 
complicates or even prevents a practical implementation of the system.
%prevents it from being effectively implemented on-chip. 
This issue can be resolved by introducing another resonator into the system (resonator $R_1$ in Fig.\ref{fig:fig3}a) operating at or near resonance 
\cite{xue2015-1,miller2015} 
and acting as a ``pre-amplifier'' between the input waveguide and the nonlinear resonator~$R_2$. To demonstrate the properties of the double resonator system, we first consider the extra resonator $R_1$ to be made of a linear optical material, with the case of both resonators made of the same nonlinear medium described in the following section.

\begin{figure}[htbp]
\centering
\includegraphics[width=1.0 \columnwidth]{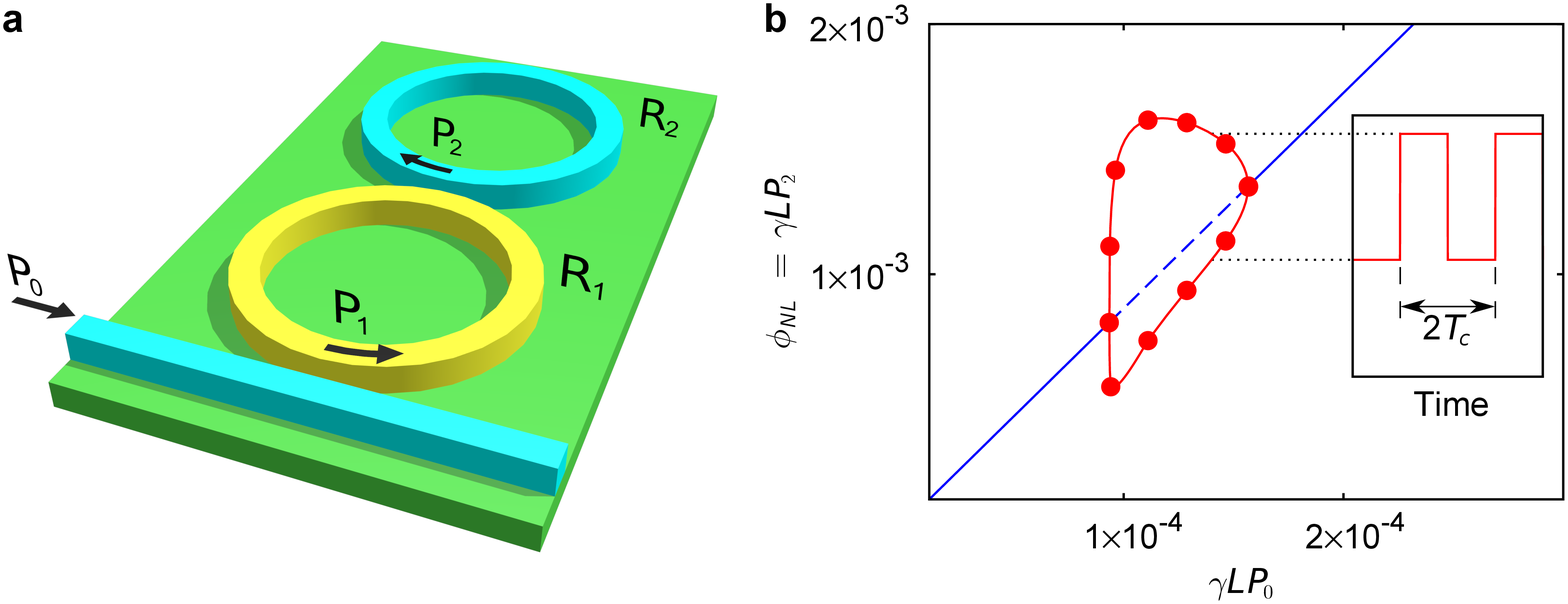}
\caption{\label{fig:fig3} (a) Schematics of the double resonator system: $R_1$ -- linear microring resonator, $R_2$ -- nonlinear. (b) Bifurcation diagram of the first period-1 instability region for the states of the power $P_2$ inside $R_2$ at different values of the input power $P_0$. 
Fixed points of period-1 and period-2 form the blue line and the red closed loop with circle markers respectively.
Solid lines represent stable fixed points, dashed -- unstable.
Inset: waveforms inside $R_2$ at the scaled input power $\gamma{L}{P_0} = 1.4\times10^{-4}$ at zero GVD. 
The simulation data were obtained for resonators with ring radii $R=10\ \mathrm{{\mu}{m}}$, $\kappa_1^2 = 0.01$, $\kappa_2^2 = 0.1$, $\phi_1 = 0$, $\phi_2 = \pi - 1.7\times10^{-3}$, $\alpha=0.7\ \mathrm{dB/cm}$, and TPA in $R_2$ as in silicon near $\lambda=1550\ \mathrm{nm}$.}
\end{figure}

Similarly to the single resonator system, we obtain the nonlinear map:
\begin{eqnarray}
b_2^{(n+1)} &=& \tau_2 a_2^{(n)} \exp(i\phi_2) + 
\tau_1 \kappa_2 a_1^{(n)} \exp(i\phi_1)\nonumber\\
&&+
\kappa_1 \kappa_2 c,
\label{eq:LR1NLR2MapEq1}
\end{eqnarray}
\begin{eqnarray}
a_2^{(n)} &=& b_2^{(n)}\frac{\exp\left(-\frac{\alpha}{2}L\right)}{\sqrt{1+2r\gamma\tilde{L}{|b_2^{(n)}|}^2}}\nonumber\\
&&\times
\exp\left[{i\log\left(1+2r\gamma\tilde{L}{|b_2^{(n)}|}^2\right)}/(2r)\right],
\label{eq:LR1NLR2MapEq2}
\end{eqnarray}
\begin{equation}
a_1^{(n)} = d_2^{(n)} \exp\left(-\frac{\alpha}{2}L\right),
\label{eq:LR1NLR2MapEq3}
\end{equation}
\begin{equation}
d_2^{(n)} = \frac{\tau_2^* b_2^{(n)} - a_2^{(n-1)}}{\kappa_2},
\label{eq:LR1NLR2MapEq4}
\end{equation}
where we assumed the equal roundtrip time $T_c = L/v_g$ in both resonators of equal circumference $L$ for simplicity of the analysis, $\phi_1$ and $\phi_2$ are the linear phase detunings of the pump frequency from the closest resonant modes in the cavities $R_1$ and $R_2$ correspondingly, $\tau_1$ ($\kappa_1$) and $\tau_2$ ($\kappa_2$) are the amplitude transmission (coupling) coefficients for the coupler between the waveguide and $R_1$ and between $R_1$ and $R_2$ respectively, $\gamma$ and $r$ are the nonlinearity and the TPA coefficients correspondingly of the nonlinear resonator $R_2$.

As before, the stability analysis reveals multiple regions of period-1 stable and unstable states and confirms that the system has a bifurcation from period-1 to period-2 stable state when the resonator $R_2$ is tuned into the vicinity of anti-resonance (Fig.\ref{fig:fig3}b), as in the case of the single resonator, whereas the resonator $R_1$ is tuned into the near-resonance and provides the required power upconversion between the waveguide and the resonator $R_2$. 
More than that, the stability analysis also shows that the double resonator system is more unstable than the considered earlier single resonator system, which is beneficial for frequency comb generation. 
Specifically, adding an extra resonator widens the instability regions and lowers the threshold input power as well as the intracavity powers: 
%Thus, in the overcoupled resonator approximation (which is the case for the system with parameters shown in Fig.3), $\phi_{NL}^{Th} \approx \gamma{L}{P_2^{Th}} \sim \kappa_1^2\kappa_2^2$ with the optimally tuned linear phase $\phi_2 \sim \pi -\kappa_1^2\kappa_2^2$ and resonator $R_1$ in near-resonance (compare with $\phi_{NL}^{Th} \approx \gamma{L}{P^{Th}} \sim \kappa^2$ in the single resonator system), $\gamma{L}{P_1^{Th}} \sim \kappa_2^{-2}\gamma{L}{P_2^{Th}} \sim \kappa_1^2$, and $\gamma{L}{P_0^{Th}} \sim \kappa_1^{2}\gamma{L}{P_1^{Th}} \sim \kappa_1^4$ (against $\gamma{L}{P_0^{Th}} = {\rm const} \approx 2.5$ in the single resonator system).
% As before, results of simulation show that increasing linear losses over the critical coupling point gradually increases the threshold powers.
$\phi_{NL}^{Th} \equiv \gamma{L}{P_2^{Th}} \sim \kappa_1^2\kappa_2^2$ 
(against $\phi_{NL}^{Th} \equiv \gamma{L}{P^{Th}} \sim \kappa^2$ in the single resonator system), 
with the optimally tuned linear phase detuning $\phi_2 \sim \pi -\kappa_1^2\kappa_2^2$, resonator $R_1$ in near-resonance regime, and negligible linear loss approximation as before; 
$\gamma{L}{P_1^{Th}} \sim \kappa_2^{-2}\gamma{L}{P_2^{Th}} \sim \kappa_1^2$ and $\gamma{L}{P_0^{Th}} \sim \kappa_1^{2}\gamma{L}{P_1^{Th}} \sim \kappa_1^4$ (against $\gamma{L}{P_0^{Th}} = {\rm const} \approx 2.5$ in the single resonator system).
Therefore, in contrast to the case of the single resonator, the coupling between resonator $R_1$ and the waveguide in the suggested double resonator system can be designed so that the threshold input power is orders of magnitude lower ($\gamma{L}{P_0^{Th}} \approx 1\times10^{-4}$ in Fig.\ref{fig:fig3}b), which solves the high threshold input power problem.

%One could argue, that although such design could provide frequency comb generation at low input powers theoretically, it would still be difficult to implement it from a practical point of view, namely, to fabricate and couple two resonators made from different materials (resonator $R_1$ made from a highly linear material, such as silica, and $R_2$ -- from a nonlinear material, such as silicon) on the same substrate. However, as it was shown above, a linear resonator in the double resonator system acts not only as a pre-amplifier, but also lowers the minimum threshold intracavity powers by adding instability to the system. Thus, for the linear resonator we have $\gamma{L}{P_1^{Th}} \sim \kappa_1^2$ and, therefore, for $\kappa_1^2 \ll 1$ we obtain $\gamma{L}{P_1^{Th}} \ll 1$ -- the necessary condition for the multi-photon absorption in the resonator $R_1$ to be negligibly small, if $R_1$ was in fact a nonlinear resonator, the same as $R_2$. The last argument gives a reason to believe that at sufficiently low intracavity powers the linear resonator $R_1$ could be replaced with a nonlinear one in the double resonator system without any significant changes in the power distribution in the system, given that $R_1$ is still kept in the near-resonance regime. In addition, adding more nonlinearity into the system could even further lower the threshold powers, as we demonstrate below.

However, even though such double resonator design could provide frequency comb generation at low input and intracavity powers theoretically, from a practical point of view it would be difficult to fabricate and couple two resonators made from different materials ($R_1$ -- from a highly linear material, such as silica; $R_2$ -- from a nonlinear material, such as silicon) on the same substrate. In the next section we show that the double resonator system still supports frequency comb generation at low powers even when both resonators are made from the same nonlinear material.

\subsection{Nonlinear double resonator system}
%% Model & Map
Fabricating both resonators from the same nonlinear material would be easier for manufacturing reasons than making them from two different media, and, as in the case of silicon, would also be compatible with the standard silicon-based fabrication process. Since the double resonator system with a linear resonator $R_1$ is capable of frequency comb generation at low input as well as low intracavity powers, $R_1$ could be replaced with a nonlinear resonator equivalent to $R_2$, as shown in Fig.\ref{fig:fig4}a.

\begin{figure}[htbp]
\centering
\includegraphics[width=1.0 \columnwidth]{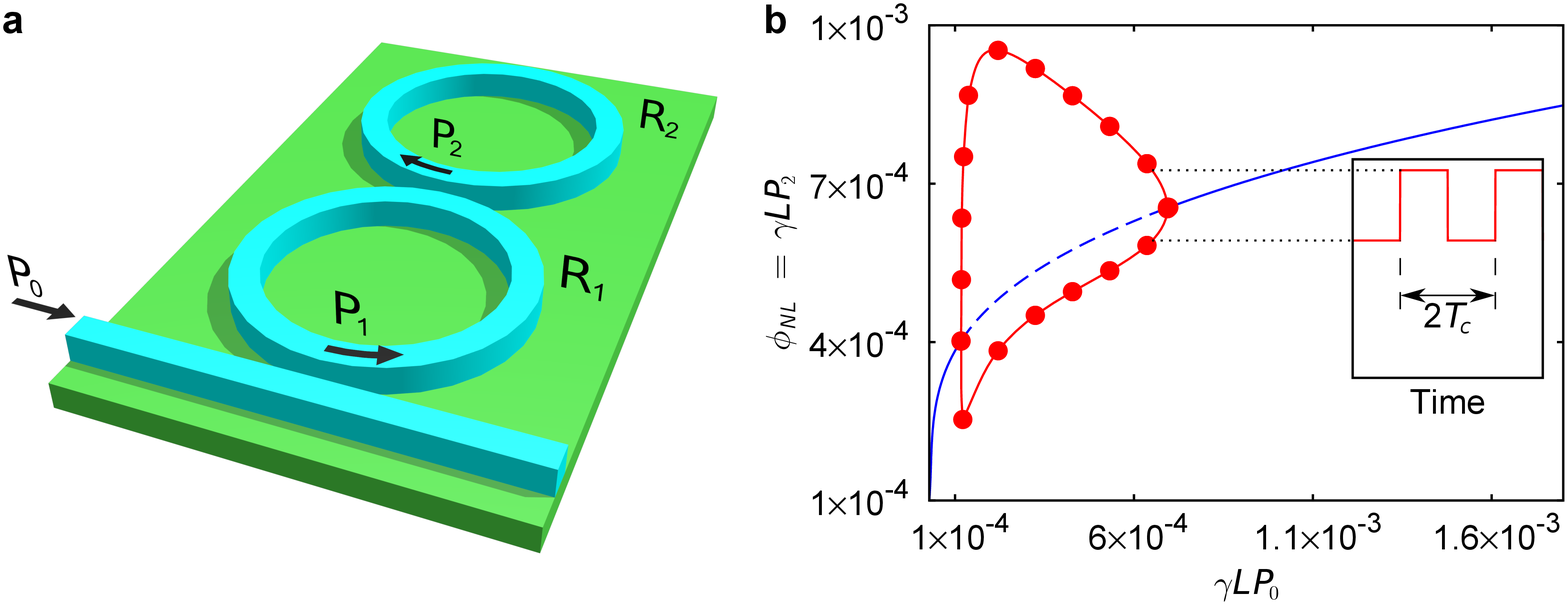}
\caption{\label{fig:fig4} (a) Schematics of the nonlinear double resonator system: $R_1$ and $R_2$ -- nonlinear microring resonators. (b) Bifurcation diagram of the first period-1 instability region for the states of the power $P_2$ inside $R_2$ at different values of the scaled input power $P_0$.
Fixed points of period-1 and period-2 form the blue line and the red closed loop with circle markers respectively.
Solid lines represent stable fixed points, dashed -- unstable.
Inset: waveforms inside $R_2$ at the scaled input power $\gamma{L}{P_0} = 6.5\times10^{-4}$ at zero GVD.
The simulation data were obtained for resonators with ring radii $R=10\ \mathrm{{\mu}{m}}$, $\kappa_1^2 = 0.01$, $\kappa_2^2 = 0.1$, $\phi_1 = -1\times10^{-2}$, $\phi_2 = \pi - 1.4\times10^{-3}$, $\alpha=0.7\ \mathrm{dB/cm}$, and TPA as in silicon near $\lambda=1550\ \mathrm{nm}$.
For the silicon microring waveguide crossection area of $450\ \mathrm{nm} \times 220\ \mathrm{nm}$,
the value of $\gamma{L}{P} = 10^{-4}$ corresponds to $\approx 10\ \mathrm{mW}$ of power.}
\end{figure}

For the nonlinear double resonator system (Fig.\ref{fig:fig4}a) we obtain the nonlinear map:
\begin{eqnarray}
b_2^{(n+1)} &=& \tau_2 a_2^{(n)} \exp(i\phi_2) + 
\tau_1 \kappa_2 a_1^{(n)} \exp(i\phi_1)\nonumber\\
&&+
\kappa_1 \kappa_2 c,
\label{eq:NLR1NLR2MapEq1}
\end{eqnarray}
\begin{eqnarray}
a_2^{(n)} &=& b_2^{(n)}\frac{\exp\left(-\frac{\alpha}{2}L\right)}{\sqrt{1+2r\gamma\tilde{L}{|b_2^{(n)}|}^2}}\nonumber\\
&&\times
\exp\left[{i\log\left(1+2r\gamma\tilde{L}{|b_2^{(n)}|}^2\right)}/(2r)\right],
\label{eq:NLR1NLR2MapEq2}
\end{eqnarray}
\begin{eqnarray}
a_1^{(n)} &=& d_2^{(n)}\frac{\exp\left(-\frac{\alpha}{2}L\right)}{\sqrt{1+2r\gamma\tilde{L}{|d_2^{(n)}|}^2}}\nonumber\\
&&\times
\exp\left[{i\log\left(1+2r\gamma\tilde{L}{|d_2^{(n)}|}^2\right)}/(2r)\right],
\label{eq:NLR1NLR2MapEq3}
\end{eqnarray}
\begin{equation}
d_2^{(n)} = \frac{\tau_2^* b_2^{(n)} - a_2^{(n-1)}}{\kappa_2},
\label{eq:NLR1NLR2MapEq4}
\end{equation}
where we assumed the two resonators to be identical, each with the roundtrip time $T_c = L/v_g$, circumference $L$, nonlinearity $\gamma$ and the TPA $r$ coefficients; $\phi_{1(2)}$, $\tau_{1(2)}$, and $\kappa_{1(2)}$ are defined the same way as before.

As seen from Fig.\ref{fig:fig4}b, frequency comb generation is possible when the resonator $R_2$ is tuned into the vicinity of anti-resonance and the resonator $R_1$ is tuned into the near-resonance, just as in the previously considered case of the linear ``pre-amplifier'' $R_1$.
The intracavity power in $R_1$ is sufficiently low ($\gamma{L}{P_1^{Th}} \sim 10^{-2}$), so that the multi-photon absorption processes are negligible. 
Note that the value of the linear phase detuning in $R_1$ is picked in such a way ($\phi_1 = -1\times10^{-2}$), 
that the resonator operates close to its resonance, but does not enter the bistability regime.
As it could have been expected, with introducing nonlinearity in the resonator $R_1$, 
the overall system becomes more unstable, which can be observed as broadening of 
the period-1 instability region -- which makes the frequency comb generation regime 
accessible at wider range of input powers 
($\gamma{L}{P_0} \approx 1\times10^{-4} \dots 7\times10^{-4}$).

\section{Discussion}

Now that we introduced a new (anti-resonant) mechanism of frequency comb generation, 
one of the key questions to answer is how it compares with the conventional (resonant) method \cite{kippenberg2004,ferdous2011,herr2012}.
First of all, as it was mentioned before, the anti-resonant mechanism has a specific spectral signature: the first harmonic is generated at $f_1^{(2)} = \frac{FSR}{2} = \frac{1}{2 T_c}$ away from the pump frequency for a period-$2$ state ($f_1^{(n)} = \frac{FSR}{n} = \frac{1}{n T_c}$ for a period-$n$ state), while in the resonant approach all harmonics are formed at multiple or single FSR away from the pump.
Secondly, if the comb in the suggested mechanism arises as a consequence of the period-doubling bifurcation, comb generation in the conventional method is connected to the existence of MI gain and cavity solitons.
%temporal pattern formation (intensity switching with each roundtrip), comb generation in the conventional method is connected to the formation of spatial patterns, such as cavity solitons and Turing patterns, along the resonator circumference \cite{coillet2013}.
% Generally, in reality, the pattern would be both spatial and temporal (spatiotemporal) (citation: coillet, LL, others?). Whichever dominates, determines the mechanism.
Thirdly, the group velocity dispersion is the key factor for pattern formation in the resonant approach, while the comb generation with the anti-resonant mechanism relies exclusively on the first-order dispersion (group velocity) for pattern formation and non-zero normal GVD only narrows the spectrum.
%The key role of GVD for pattern formation in the resonant case can be seen from the bifurcation diagram for zero-GVD case (Fig.\ref{fig:fig1}b): the system exhibits period-doubling instabilities near the cavity anti-resonances, which lead to the temporal pattern formation and frequency comb generation, while the system dynamics in the resonant regions is limited by bistable behaviour. Non-zero GVD enables interaction between adjacent infinitesimally short pulse slices propagating along the resonator circumference and thus introducing a new (spatial) dimension into the nonlinear dynamics of the system, as well as spatial patterns and the resonant regime of frequency comb generation.

\begin{figure}[htbp]
\centering
\includegraphics[width=.60\columnwidth]{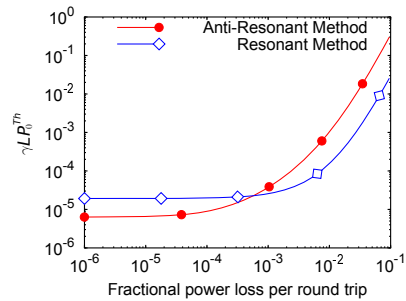}
\caption{\label{fig:fig5} Input power threshold for frequency comb generation in the resonant (blue curve with open diamonds) and anti-resonant (red curve with solid circles) regimes at different levels of linear losses and without nonlinear losses.
The simulation data were obtained in the resonant regime for the resonator with $\kappa^2 = 0.01$;
in the anti-resonant regime -- for the nonlinear double resonator system with $\kappa_1^2 = 0.01$, $\kappa_2^2 = 0.1$.
Given the silicon nitride microring waveguide crossection area of $1.3\ \mathrm{{\mu}{m}} \times 600\ \mathrm{nm}$ and the ring radius of $60\ \mathrm{{\mu}{m}}$ \cite{xue2015-1},
the value of $\gamma{L}{P} = 10^{-5}$ corresponds to $\approx 20\ \mathrm{mW}$ of power.}
\end{figure}

The existence of a power threshold for parametric oscillation in the resonant regime with anomalous dispersion can be interpreted as a balance between the parametric gain of the NLS equation and the losses in the cavity \cite{kippenberg2004,matsko2005}.
The threshold power for the anti-resonant frequency comb generation is determined by the first period-doubling bifurcation.
Fig.\ref{fig:fig5} illustrates the difference in threshold powers for frequency comb generation between the resonant and anti-resonant regimes. For the given parameters, the anti-resonant method demonstrates the lowest threshold power at low linear losses, while the resonant one -- at high losses.
%However, note that Fig.\ref{fig:fig5} shows the case of zero nonlinear losses: frequency comb generation in the resonant regime 
%requires the resonator to be made from a material without nonlinear losses at the frequency range of operation \cite{lau2015}, making the anti-resonant method a possible solution to the problem of frequency comb generation in the presence of nonlinear losses.

\section{The origin of period-doubling instability}

The comb generation in a nonlinear resonator is a special example of pattern formation 
in nonlinear systems, and as such follows the general rules of nonlinear dynamics \cite{feigenbaum1983,devaney1992}.
The onset of the frequency comb formation is related to the loss of stability of the ``original''
CW state.
This loss of stability proceeds via one of the standard
bifurcation processes \cite{devaney1992}, such as the period-doubling bifurcation \cite{feigenbaum1983}.
To understand the origin of the comb formation in our system, it is therefore necessary
to identify the point where nonlinearity will induce this bifurcation.

A CW state of a cavity is stable if perturbations (deviations from the CW amplitude value)
diminish with every roundtrip in the cavity.

If deviations change sign after every roundtrip while conserving their amplitude, %without attenuation, 
we observe a possible point of the onset of the period-doubling instability: 
after the first roundtrip intracavity intensity
deviates from its CW value, and the next roundtrip it returns to its original value.
In contrast to the CW mode with a single value of the intracavity intensity -- period-1 state,
the new regime has a period of two roundtrips with two unique values of the intracavity intensity
(period-2 state).
Thus, the corresponding %amplification 
coefficient for deviations from the CW-state has a special meaning
to the stability of the cavity and, when equal to $-1$ 
(change of sign without change in deviation intensity), marks the onset 
of the period-doubling instability. In literature \cite{devaney1992}, such amplification coefficient is known as 
an eigenvalue of the monodromy matrix for perturbations, and the onset point -- as the period-doubling
bifurcation point \cite{feigenbaum1983}.

In the purely linear and lossless cavity, 
it is the anti-resonance and only the anti-resonance point,
that in the limit of decoupled cavity ($\tau \rightarrow 1$) 
has a real eigenvalue approaching $-1$. % $\lambda \rightarrow -1$ .
Thus, with finite nonlinearity added to the system, a period-doubling instability region forms 
in the system phase-space around this marginally stable point, 
and inside of this region frequency comb generation can be observed.
%Below we show how this follows analytically.

Below, we analytically demonstrate this behaviour
(eigenvalue $\lambda \rightarrow -1$ at anti-resonance as $\tau \rightarrow 1$).
We study the system stability by analyzing the robustness of its CW-state to perturbations.
Since the physical mechanism behind the onset of period-doubling instability 
is the same in the single cavity and the double resonator system, 
for reasons of clarity we will duscuss here the simplest case of a single cavity without losses.
The map describing the dynamics of the lossless linear coupled cavity is a special case
of the map defined by Eqs. (\ref{eq:NLRMapEq1},\ref{eq:NLRMapEq2}) for $\gamma = 0$,
and has a CW-solution in the steady-state, when $b^{(n)} = b = \mathrm{const}$:
\begin{eqnarray}
b = \kappa c + \tau b \exp\left(i\phi\right),
\label{eq:LRCW}
\end{eqnarray}
To check the stability of this CW-solution we perturb 
the steady state slightly such that $b^{(n)} = b + u^{(n)}$. 
From Eqs. (\ref{eq:NLRMapEq1},\ref{eq:NLRMapEq2}) for $\gamma = 0$ we obtain
\begin{eqnarray}
b + u^{(n+1)} = 
\kappa c + \tau (b + u^{(n)}) \exp\left(i\phi\right)
\label{eq:LRCWpMap1}
\end{eqnarray}
After subtracting Eq. (\ref{eq:LRCW}) from Eq. (\ref{eq:LRCWpMap1}), we arrive to the linear map for perturbations
\begin{eqnarray}
u^{(n+1)} = \lambda u^{(n)}
\label{eq:LRCWpMap2}
\end{eqnarray}
with eigenvalue
\begin{eqnarray}
\lambda = \tau \exp{\left({i}\phi\right)}
\label{eq:LRCWpMapEig}.
\end{eqnarray}

From Eq. (\ref{eq:LRCWpMapEig}) it follows that only at the antiresonance, when $\phi = \pi$,
the eigenvalue is real and negative, 
and in the limit of decoupled cavity ($\tau \rightarrow 1$)
it assumes the value of $-1$.
With nonlinearity continuously added to the system, an instability region grows around 
this point of marginal stabilility -- a region of period-doubling instability, 
so that period-doubling and the related frequency comb generation
can be observed at finite coupling, detuning from the cavity anti-resonance, and input power.

%% Conclusion
\section{Conclusion}
In conclusion, we have presented an alternative, anti-resonant, approach to nonlinear optical processes at low powers, and demonstrated its application to low-power optical frequency comb generation in a silicon chip.
Theoretical analysis and simulation results showed that the new mechanism is capable of operating at low intracavity and input powers, and is not suppressed in the presence of two-photon absorption losses in the materials such as silicon at the telecom wavelength.
%Theoretical analysis and simulation results showed that the new mechanism is capable of operating at low intracavity and input powers, and is not suppressed in the materials with nonlinear losses, such as silicon at the telecom wavelength.%1550 nm.

%% Acknowledgements
\begin{acknowledgments}
The research was supported by 
the Air Force Office of Scientific Research (AFOSR) under grant No. FA9550-12-1-0236
and the Gordon and Betty Moore Foundation.
The authors thank Prof. A. M. Weiner for the helpful discussions and comments on the article.
\end{acknowledgments}

%\bibliography{freqcombs-zot-bbt}

\end{document}